\documentclass[10pt,letterpaper]{article}
\usepackage{opex3}

\usepackage{graphicx}% Include figure files
\usepackage{dcolumn}% Align table columns on decimal point
\usepackage{bm}% bold math
\usepackage{quantum}

\def\r{\textbf{r}}
\def\domega{\Delta \omega (k)}
\def\Ef{\overrightarrow{\mathbf{E}}}
\def\Ew{\overrightarrow{\mathbf{E_w}}}
\def\Af{\overrightarrow{\mathbf{A}}}
\def\Bk{\overrightarrow{\mathbf{B_k}}}

\begin{document}

\title{Coupled mode theory for photonic crystal cavity-waveguide interaction\\}
\author{Edo Waks and Jelena Vuckovic}

\address{E.L. Ginzton Laboratories \\
Stanford University  \\
Stanford, CA 94305 }

\email{edo@stanford.edu}

\begin{abstract}
We derive a coupled mode theory for the interaction of an optical
cavity with a waveguide that includes waveguide dispersion. The
theory can be applied to photonic crystal cavity waveguide
structures. We derive an analytical solution to the add and drop
spectra arising from such interactions in the limit of linear
dispersion. In this limit, the spectra can accurately predict the
cold cavity quality factor (Q) when the interaction is weak. We
numerically solve the coupled mode equations for the case of a
cavity interacting with the band edge of a periodic waveguide,
where linear dispersion is no longer a good approximation. In this
regime, the density of states can distort the add and drop
spectra.  This distortion can lead to more than an order of
magnitude overestimation of the cavity Q.
\end{abstract}

\section{Introduction}
\label{Sec1:Introduction}

High-Q photonic crystal (PC) resonators have recently become a
subject of great interest.  Such cavities have important
applications for low threshold lasers, high finesse filters, as
well as experiments in cavity quantum electrodynamics
(CQED)~\cite{LocarYoshie2002,AkahaneMochizuki2003,VuckovicYamamoto2003}.
One advantage of using PC resonators is that they can be easily
integrated with PC based waveguide structures.  This is important
for integrated optical processing.

The interaction of a cavity resonator with a waveguide system has
been theoretically studied previously in~\cite{ManolatouKhan99}.
That work considered waveguides with continuous translation
symmetry and ignored waveguide dispersion. Such approximations are
often good for optical fiber waveguides, but do not necessarily
apply to PC based waveguides. These waveguides are periodic in the
direction of propagation, and hence exhibit discrete instead of
continuous translation symmetry. Because of the discrete
translation symmetry, the modes of the waveguide are no longer
simple travelling waves.  Instead, they take on the form of Bloch
states. Another consequence of the waveguide periodicity is that
it features an energy stop-band.  At the edge of the stop band,
the group velocity goes to zero and the dispersion becomes
important in characterizing the interaction between the cavity and
waveguide.  The porperties of the interaction near the band edge
are particularly important when using photonic crystal cavities
formed by single or multiple hole row defects.  The modes of these
type of cavities are primarily coupled to waveguide modes near the
band edge.

The main goal of this paper is to investigate the interaction of
photonic crystal based cavities and waveguides using coupled mode
theory as in~\cite{ManolatouKhan99}.  However, in order to apply
coupled mode theory, we must properly incorporate dispersion,
which plays an important role in photonic crystal waveguides.  One
of the main results of this paper is a set of coupled mode
equations that include dispersion and properly handle the Bloch
mode structure of the waveguide modes. Once this is derived, we
can apply the theory to realistic photonic crystal based systems.

We first derive the equations of motion for the coupled mode
system.  After deriving these equations, we solve the system
analytically for the special case where the waveguide dispersion
relation can be approximated by linear dispersion. Expressions for
the add filter and drop filter spectra are explicitly given. When
the dispersion relation can no longer be approximated as linear,
as in the case of a periodic waveguide near the stop band, an
analytical solution becomes too difficult to derive. Instead, we
simulate the equations of motion numerically to find the solution.

Our simulations focus on the drop filtering spectrum of the
system.  Drop filtering is an important operation to analyze
because it is often used to measure the cavity quality factor
(Q)~\cite{AkahaneAsano2003}.  To properly interpret such results,
it is important to understand the limits under which these
measurements can be used to infer Q.  We investigate two cases of
waveguides that feature stop-bands. The first is a waveguide with
weak periodicity in the direction of propagation. In this limit
the waveguide dispersion relation is parabolic. Although weak
periodicity is rarely a good approximation for photonic crystals,
it provides a good toy model of a structure with a stop-band, and
elucidates much of the physical intuition about the problem.  In
the second case, we apply the coupled mode theory to the realistic
case of a row-defect photonic-crystal waveguide coupled to a
three-hole defect cavity. The modes of the cavity and waveguide,
along with the waveguide dispersion relation, are first calculated
using FDTD simulation. These simulations are used to calculate the
coupling coefficients which enter into the coupled mode theory.
The system is then simulated, giving what we believe to be an
accurate analysis of a real experiment using such structures.  In
both cases, we show that when the cavity is resonant near the
stop-band, the cavity Q can be overestimated by more than an order
of magnitude. This is because the interaction of the cavity with
the waveguide is determined by both the cavity spectral function,
as well as the waveguide density of states. Near the band edge,
the density of states diverges leading to a sharp spectral feature
that is unrelated to cavity properties.

\section{Coupled Mode Theory}
\label{Sec2:Coupled Mode Theory}

We begin the derivation of the coupled mode equations with the
wave equation
\begin{equation}\label{eq:WaveEq}
\nabla\times\nabla\times \Ef +
\frac{\epsilon(\overrightarrow{r})}{c^2}
\frac{\partial^2\Ef}{\partial t^2}=0
\end{equation}
where $\epsilon(\r)$ is the relative dielectric constant, and $c$
is he speed of light in vacuum.  We define $\epsilon_c$ as the
relative dielectric constant for the cavity, $\epsilon_w$ as the
dielectric constant for the waveguide, and $\epsilon_t$ for the
coupled system. We assume the waveguide dielectric constant to be
periodic.  Thus, the solutions to Eq.~\ref{eq:WaveEq} with
$\epsilon=\epsilon_w$, denoted $\Ew$, must satisfy the Bloch
theorem, and hence take on the form
  \begin{equation}
    \Ew = \Bk (\r)e^{i(\omega (k)t - kz)}
  \end{equation}
where $\Bk$ are Bloch states that have the same periodicity as
$\epsilon_w$, $k$ is the crystal momentum, and $z$ the direction
of propagation in the waveguide.  The cavity mode, which is the
solution to Eq.~\ref{eq:WaveEq} with $\epsilon=\epsilon_c$ as the
index, is defined as $\Af(\r)$.

The dynamics of the coupled system are determined by setting
$\epsilon=\epsilon_t$ in Eq.~\ref{eq:WaveEq}.  Using the standard
arguments of coupled mode theory~\cite{Yariv}, we assume the
solution of the coupled system to take on the form
  \begin{equation}
    \Ef = a(t) \Af(\r) e^{i\omega_c t} + \int dk \Bk(\r)
    e^{i\omega(k) t} \left[ b(k,t) e^{-ikz} + c(k,t) e^{ikz}
    \right]
  \end{equation}
where $a(t)$ is the slowly varying component of the cavity, and
$b(k,t)$  and $c(k,t)$ are slowly varying components of the
forward and backward propagating Bloch states respectively.
Plugging the above solution back into Eq.~\ref{eq:WaveEq}, we
derive the coupled mode equations
  \begin{eqnarray} \label{eq:Diffeq1}
    \frac{da}{dt} & = & - i \int dk \frac{\omega^2(k)}{\omega_c}
    e^{i \domega t}\left[ b(k,t) \kappa_{ba}(k) + c(k,t)\kappa_{ba}(-k)\right]
     - \lambda a +P_c e^{i(\omega_p - \omega (k))t}\\ \label{eq:Diffeq2}
    \frac{db(k)}{dt} & = & -i
    \frac{\omega_c^2\kappa_{ab}(k)}{\omega(k)}ae^{-i \domega t} +
    P_w (k) e^{i(\omega_p - \omega (k))t} -\eta b(k)\\ \label{eq:Diffeq3}
    \frac{dc(k)}{dt} & = & -i
    \frac{\omega_c^2\kappa_{ab}(-k)}{\omega(k)}ae^{-i\domega t} +
    P_w (-k) e^{i(\omega_p - \omega (k))t} -\eta c(k)
  \end{eqnarray}
In the above equations, $\lambda$ is a phenomenological decay
constant which is added to account for the finite lifetime of the
cavity resulting from mechanisms other than cavity-waveguide
coupling. $P_c$ and $P_w (k)$ are external driving terms that can
potentially drive the cavity or waveguide at a frequency
$\omega_p$.  The damping term $\eta$ is also included to give the
waveguide modes a finite lifetime.  In the analytical calculations
we take the limit $\eta \to 0$.  In the numerical simulations,
however, we set this damping term to a very small value in order
to have a well defined steady state solution.  The coupling
constants are given by
  \begin{eqnarray}
    \kappa_{ba} (k) & = &
    \frac{\int d\r \frac{\Delta\epsilon_w(\r )}{c^2}
    e^{-ikz}\Bk \cdot \Af^*}
    {\int d\r \frac{2 \epsilon_t(\r)}{c^2} |\Af|^2}\label{eq:Kappa_ab} \\
    \kappa_{ab} (k) & = &
    \int d\r \frac{\Delta\epsilon_c(\r )}{c^2}
    e^{-ikz}\Af\cdot \Bk^* \label{eq:Kappa_ba}
  \end{eqnarray}
where $\Delta \epsilon_{c,w} = \epsilon_t - \epsilon_{c,w}$.

\section{Linear dispersion}

The solution to the above set of coupled equations strongly
depends on the waveguide dispersion relation, which relates
$\omega(k)$ to $k$. For some systems, we can assume that this
relation is linear, taking on the form
  \begin{equation}
    \omega(k) = \omega_0 + V_g (k-k_0)
  \end{equation}
where $V_g$ is the group velocity.  When this linearized
approximation is valid, an analytical solution can be derived for
Eq.~\ref{eq:Diffeq1}-\ref{eq:Diffeq3}.  This solution is most
easily obtained using the method of Laplace transforms.  We take
the Lapace transform in time of Eq.~\ref{eq:Diffeq2} and
Eq.~\ref{eq:Diffeq3} and plug into Eq.~\ref{eq:Diffeq1}.  We make
the additional approximation
  \begin{equation}
    \frac{1}{(\omega_{c,p} - \omega(k))+is} \approx P\left[ \frac{1}{\omega_{c,p} - \omega(k)} \right] + i
    \pi \delta(\omega_{c,p} - \omega(k))
  \end{equation}
where $P$ represents the Cauchy principal value of the expression.
This leads to
  \begin{equation}  \label{eq:LaplaceEq}
    a(s) = \frac{1}{s+\lambda + \Gamma + i\delta \omega}\left( a_0 +
    \frac{P_c + J}{(s- i (\omega_p - \omega_c))} \right)
  \end{equation}
where, $a_0$ is the initial cavity field, and the other constants
are given by
  \begin{eqnarray*}
    \Gamma & = & \frac{2\pi Re \left\{\kappa_{ab}(k(\omega_c))
    \kappa_{ba}(k(\omega_c))\right\}}{V_g} \\
    \delta \omega & = & P\left( \int dk \frac{2\omega(k) Re \left\{\kappa_{ab}(k(\omega_c))
    \kappa_{ba}(k(\omega_c))\right\} }{V_g(\omega(k) - \omega_c)} \right) \\
    J & = & P\left( \int dk \frac{2\omega (k) \kappa_{ab}(k)}{\omega(k)
    - \omega_c} \right) - i \frac{\pi \omega_c (\kappa_{ab}(k(\omega_c))P_w(k(\omega_c)) +
    \kappa_{ab}(-k(\omega_c))P_w(-k(\omega_c))}{V_g}
  \end{eqnarray*}
The above expressions also assume that $A(r)$ is a real function,
so that $\gamma(k)=\gamma(-k)$.

Consider first the simple example of a ring-down experiment with
no external sources, meaning $P_w=P_c=0$. The cavity is assumed to
contain an initial field $a(0)$ at time $0$. The solution of the
cavity field is obtained from the equations of motion to be
  \begin{equation} \label{eq:CavityField}
    a(t) = a(0)e^{-(\lambda + \Gamma )t}
  \end{equation}
The above solution has a simple interpretation.  The constant
$\lambda$ is the rate at which the cavity field escapes into leaky
modes, while $\Gamma$ is the rate at which the cavity field
escapes into the waveguide.  The total decay rate of the cavity
field is simply the sum of these two rates.  It is important to
note that the coupling rate into the waveguide is inversely
proportional to the group velocity.  This dependence is simply a
reflection of the increased interaction time between the cavity
and waveguide at slower group velocities.

Next consider an add filter experiment, where both cavity and
waveguide are initially empty and $P_w=0$.  One can show that the
cavity source term will drive the waveguide field to a steady
state value given by
  \begin{equation}
    |b_k(t)|^2 =  \frac{|\pi\kappa_{ab} P_c|^2}{\omega
    (k)^2} \frac{1}{(\omega_p - \omega_c + \delta \omega)^2 + (\lambda
    + \Gamma)^2}
  \end{equation}
The field features the Lorentzian line-shape expected from an
exponential decay process.  Similarly, one can derive the drop
spectrum of the waveguide by setting $P_c=0$.  In this case the
waveguide spectrum is
  \begin{equation} \label{eq:DropSpectrum}
    |b_k (t)|^2 = |P_w (k_p)|^2 \left|\left( 1 - \frac{J}{i (\omega_p - \omega_c )
    + \lambda + \Gamma}\right)\right|
  \end{equation}
which is an inverse Lorenztian.

\section{Weakly periodic waveguides}

In the linear dispersion limit there is little qualitative
difference between the results presented  above and those for the
single mode analysis in Ref.~\cite{ManolatouKhan99}.  The main
distinction is that with linear dispersion, the interaction
strength is inversely proportional to the group velocity.  But in
many cases, one cannot linearly approximate the dispersion
relation. One such case is a photonic crystal waveguide, which is
periodic in the direction of propagation and therefore features a
stop-band.  At the edge of the stop-band, the group velocity goes
to zero, at which point the dispersive properties of the waveguide
become important.

\begin{figure}
\centering\includegraphics[width=7cm]{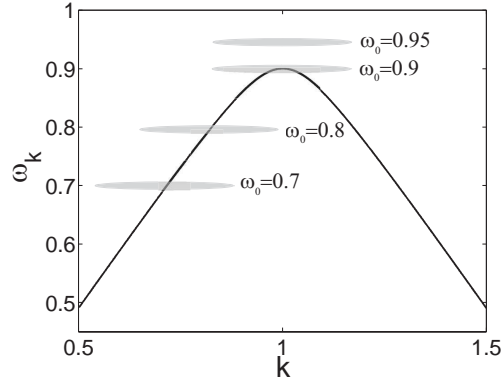}
\caption{Dispersion relation for a periodic waveguide.}
\label{fig:dispersion}
\end{figure}

Before treating the full case of photonic crystal structures, we
start with a simpler case of a waveguide with weak periodicity in
the direction of propagation.  The dispersion relation of such a
structure can be approximated as~\cite{Yariv}
  \begin{equation}\label{eq:dispersion}
     \omega_k = \frac{c}{n_{eff}} \left( \frac{\pi}{\Lambda} - \sqrt{ D^2 +
     \left(k-\frac{\pi}{\Lambda}\right)^2 } \right)
   \end{equation}
Above, $\Lambda$ is the periodicity of the lattice, $D$ is the
size of the bandgap (related to the index contrast of the
periodicity), and $n_{eff}$ is an effective index of refraction.
This dispersion relation is plotted in Fig.~\ref{fig:dispersion}.

To get an intuitive understanding of how a cavity will interact
with a waveguide featuring such a dispersion, we first note that
the cavity only couples well to waveguide modes which conserve
both energy and momentum.  Because the cavity field is confined in
both space and energy, a cavity mode can be represented as a
region in the dispersion diagram.  In this work, we consider
cavities which are spatially localized to only a few wavelengths,
but have quality factors of hundreds or more.  These cavities are
highly localized in energy, and very delocalized in momentum.  We
thus represent them as an elongated ellipse on the dispersion
diagram.  In Figure~\ref{fig:dispersion} four different cavity
resonant frequencies have been drawn. Cavities with resonant
frequencies of $0.7$ and $0.8$ lie in the nearly linear region of
the dispersion diagram. This region can be treated analytically,
as we have done above. The interaction between the cavity and
waveguide mode is primarily determined by energy conservation. If
the waveguide is initially excited, modes which are near the
cavity resonance will preferentially be scattered. The transmitted
spectrum can then be used to infer the cavity spectrum, as
Eq.~\ref{eq:DropSpectrum} indicates.

Next, consider the cavity with a resonant frequency of $0.9$,
which is right at the band edge of the waveguide.  In this case,
the interaction with the waveguide is not simply determined by
energy conservation.  The cavity scatters more strongly in regions
with higher density of states, leading to distortion of the line
shape. In this case, the transmission spectrum of the waveguide is
no longer a good representation of the cavity spectrum, and may
lead to false prediction of the cavity Q.  We verify this by
numerically simulating Eq.~\ref{eq:Diffeq1}-\ref{eq:Diffeq3} using
the dispersion relation in Eq.~\ref{eq:dispersion}.

In the simulation, the speed of light is set to  $c=1$, and the
effective index of refraction is $n_{eff}=1$.  The bandgap
constant is set to $D=0.1$. We set
$\kappa_{ab}=\kappa_{ba}=10^{-2}$, and assume that these coupling
constants are independent of $k$. This is a good approximation for
small cavities which are highly delocalized in momentum. The
cavity decay constant is set to $0.01$, which corresponds to a
cavity Q of 35 for $\omega_c=0.7$.  This value was selected
because it corresponds to a sufficiently narrow linewidth for the
simulation, but is not exceedingly narrow that it requires very
long simulation times. To simulate drop filtering we set both
waveguide and cavity to be initially empty, and pump the waveguide
modes with a pump source whose resonant frequency $\omega_p$ is
swept across the cavity resonance. We set the waveguide modes to
have a decay constant $\eta = 0.0005$, which is much smaller than
the decay of the cavity, and pump the system until a steady-state
value is reached. We then calculate the transmitted power which is
defined as
  \begin{equation}
    P_T = \int_k dk \left| b(k,t_f) \right|^2
  \end{equation}
where $t_f$ is a large enough time for all transients to decay so
that the system is in steady state.  The transmitted power is
normalized by the transmitted power of the waveguide without a
cavity.  This normalization constant is calculated by evolving the
system with $\kappa_{ab}=\kappa_{ba}=0$.

\begin{figure}
\centering\includegraphics[width=7cm]{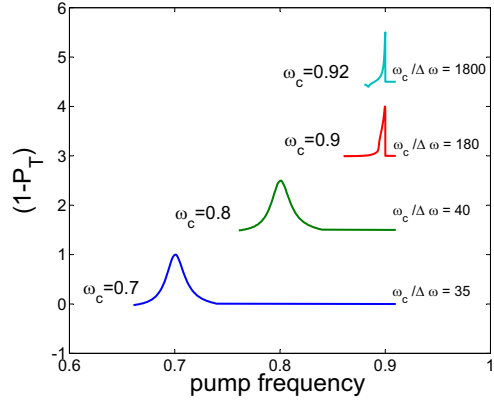}
\caption{Inverted waveguide transmission spectrum for different
cavity resonant frequencies.  Each spectrum has been normalized to
its peak value.} \label{fig:transcurveweak}
\end{figure}

The transmission as a function of pump frequency is shown in
Figure~\ref{fig:transcurveweak}.  The transmission is plotted for
a cavity resonant frequency of $0.7$, $0.8$, $0.9$, and $0.92$.
The cavities with resonant frequencies of $0.7$ and $0.8$ are in
the linear dispersion regime, so their drop spectrum is lorentzian
as predicted by  Eq.~\ref{eq:DropSpectrum}.  The linewidth of the
drop spectrum for these two frequencies has a width which
corresponds to a decay rate of $0.01$, and is therefore completely
determined by the cavity lifetime.  However, as the cavity
resonance approaches the stop-band, as for $\omega_0 = 0.9$, the
cavity spectrum significantly narrows.  This linewidth distortion
is caused by the divergence of the density of states near the band
edge.  The linewidth when $\omega_0 = 0.9$ corresponds to a
quality factor of $180$, which is significantly larger than the
cold cavity Q of $45$.  The effect is even more dramatic when
$\omega_0=0.92$, at which point the cavity resonance is completely
inside the bandgap. Despite the fact that the cavity does not
resonate with any of the waveguide modes, the extremely high
density of states near the band edge still allows the cavity to
efficiently scatter light. This results in an extremely sharp
resonance right at the band edge frequency, whose linewidth
corresponds to a Q of $1800$.

\begin{figure}
\centering\includegraphics[width=7cm]{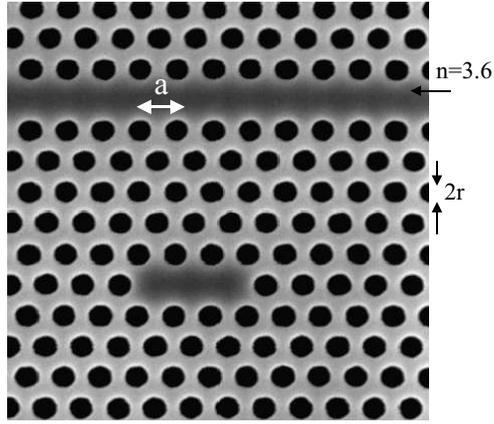} \caption{SEM
image of coupled cavity-waveguide system.} \label{fig:CoupledSEM}
\end{figure}

\section{Photonic crystal cavity-waveguide system}

\begin{figure}
\centering\includegraphics[width=7cm]{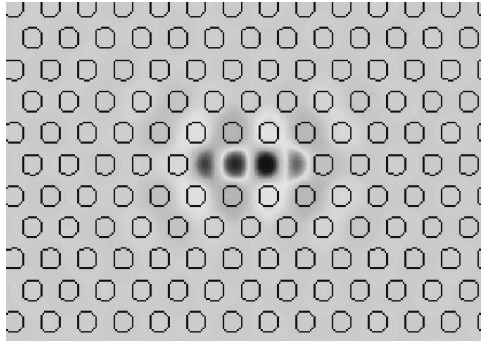} \caption{FDTD
simulation of cavity mode.  Figure shows z-component of the
magnetic field at the center of the slab.} \label{fig:CavitySim}
\end{figure}

\begin{figure}
\centering\includegraphics[width=7cm]{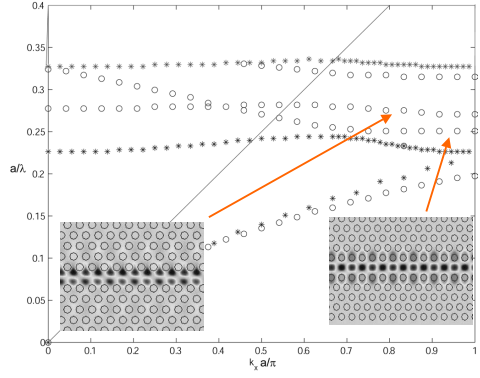}
\caption{Dispersion relation for photonic crysal waveguide.}
\label{fig:PaperDisp}
\end{figure}

We now consider the more realistic case of a photonic crystal
cavity-waveguide system.  Figure~\ref{fig:CoupledSEM} shows an SEM
image of the type of system to be analyzed.  A waveguide is formed
from a row defect in a hexagonal photonic crystal lattice with a
periodicity $a$, slab thickness $d=0.65a$, hole radius $r=0.3a$,
and refractive index $n=3.6$. The waveguide is evanescently
coupled to a cavity formed by a three hole defect.
Figure~\ref{fig:CavitySim} shows three dimensional (3D) FDTD
simulations of the cavity mode, which has a normalized resonant
frequency of $0.251$ in units of $a/\lambda$, where $\lambda$ is
the free space wavelength. Figure~\ref{fig:PaperDisp} shows the
dispersion relation of the waveguide modes, which are calculated
by the same 3D FDTD method. The stars represent the modes of the
hexagonal photonic crystal lattice.  The bandgap of the lattice
lies between the frequencies $0.23$ and $0.33$, where no
propagating modes exist. Waveguiding can only happen in this
bandgap region. The circles represent the modes of the waveguide.
Inside the bandgap these modes lie in two waveguide bands.  The
insets show the $z$ component of the magnetic field of these two
bands at the band edge, taken at the center of the slab. One of
the modes has even parity across the center of the waveguide,
while the other mode has odd parity. Looking at
Figure~\ref{fig:CavitySim}, one can see that the cavity mode has
even parity, and will therefore couple only to the even parity
Bloch state.  Thus, the odd parity mode can be neglected in the
simulations.  It is important to note that both the even and odd
modes feature a nearly flat dispersion near the band edge.

\begin{figure}
\centering\includegraphics[width=7cm]{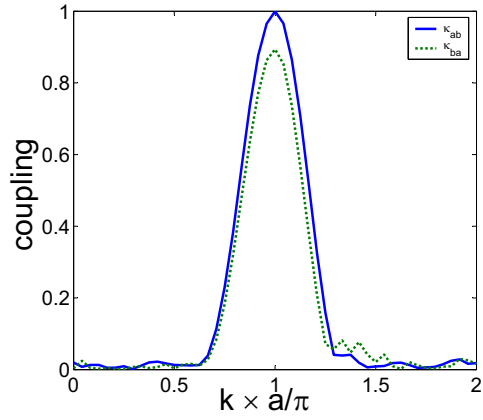}
\caption{Calculated coupling strength for cavity-waveguide
system.} \label{fig:coupling}
\end{figure}

\begin{figure}
\centering\includegraphics[width=7cm]{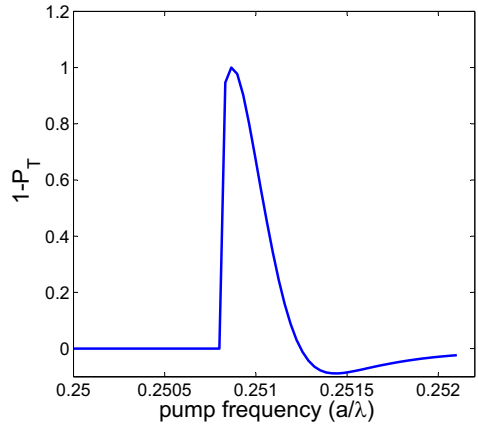}
\caption{Transmission spectrum of realistic cavity-waveguide
system.} \label{fig:phctranscurve}
\end{figure}

Next, we calculate the coupling coefficients $\kappa_{ab}$ and
$\kappa_{ba}$ using FDTD simulations as well as
equations~\ref{eq:Kappa_ab} and \ref{eq:Kappa_ba}.  The results
are shown in Figure~\ref{fig:coupling}.  The cavity is most
strongly coupled to waveguide modes near $k=\pi/a$, which is the
flattest region of the dispersion.  The calculated coupling
constants are used to simulate the waveguide transmission using
the same technique as the weak periodicity waveguide.  A three
hole defect cavity of the type shown in Fig.~\ref{fig:CoupledSEM}
has a typical Q of about 2000.  Such a high quality factor would
require extremely long calculation times to properly simulate.
Instead, we set the cavity Q 350. The drop spectrum of the cavity
is plotted in Figure~\ref{fig:PaperDisp}. From the full-width
half-max bandwidth of the cavity one finds a Q of 1300, which is
much larger than the cold cavity Q.  The width of the transmission
spectrum in Fig.~\ref{fig:phctranscurve} is limited by the
spectral resolution of the simulation.

In conclusion, we presented a coupled mode theory for
cavity-waveguide interaction which includes waveguide dispersion.
In the limit of linearly dispersion, we derived an analytical
solution for the cavity decay rate, as well as the add and drop
spectra. In this regime, the decay rate into the waveguide is
found to be inversely proportional to the group velocity.  The add
and drop spectra are also found to accurately predict the cavity
spectrum in the limit of weak interaction.  For the case of
nonlinear dispersion, we have numerically solved for the
transmission spectrum of the waveguide coupled to the cavity. We
investigated waveguides that feature a stop-band, and looked at
the behavior near the edge of the stop-band where the group
velocity vanishes. The diverging density of states near the band
edge can lead to more than an order of magnitude overestimation of
the cavity Q.  We believe these results are important in order to
better understand general cavity-waveguide interactions in most
photonic crystal systems.

This work has been supported by the MURI center for quantum
information systems (ARO/ARDA Program DAAD19-03-1-0199) and by the
Department of Central Intelligence postdoctoral fellowship
program.  The authors would also like to thank Dirk Englund for
his help with FDTD simulations, and David Fattal for assistance
with analytical coupled mode theory solutions.

\end{document}